\documentclass[11pt,a4paper]{jpconf}

\usepackage[T1]{fontenc}
\usepackage[utf8]{inputenc}
\usepackage[british]{babel}

\usepackage{color,xcolor}

\definecolor{dark-red}{rgb}{0.8, 0.0, 0.1803921568627451}
\definecolor{dark-blue}{rgb}{0.0, 0.0, 0.803921568627451}
\definecolor{dark-green}{rgb}{0.0, 0.39215686274509803, 0.0}
\definecolor{dark-orange}{rgb}{0.8, 0.4, 0.0}

\usepackage{amsmath,dsfont,mathtools,amssymb}
\usepackage{siunitx}
\usepackage{graphicx}
\usepackage{multirow,booktabs,tabularx,float}
\usepackage[font=normalsize]{subfig}
\usepackage{slashed,braket}
\usepackage{feynmp-auto}
\usepackage[sort&compress,numbers]{natbib}
\PassOptionsToPackage{hyphens}{url}\usepackage[pdftex,hidelinks,linkcolor={dark-blue}, citecolor={dark-green}, urlcolor={dark-red},bookmarksopen,linktoc=all]{hyperref}
\usepackage{enumitem} 
\usepackage[stretch=15,shrink=15]{microtype} 
\usepackage{tabto}
\usepackage{accents} 
\usepackage{tikz}
\usepackage{xspace} 
\usetikzlibrary{arrows, decorations.markings, arrows.meta}
\usepackage{pythonhighlight} 

\makeatletter
\def\l@subsubsection#1#2{}
\makeatother

\AtBeginDocument{
  \heavyrulewidth=.08em
  \lightrulewidth=.05em
  \cmidrulewidth=.03em
  \belowrulesep=.65ex
  \belowbottomsep=0pt
  \aboverulesep=.4ex
  \abovetopsep=0pt
  \cmidrulesep=\doublerulesep
  \cmidrulekern=.5em
  \defaultaddspace= .5em
  \setlength{\tabcolsep}{0.5em}
}

\setlist[itemize]{itemsep=1pt,parsep=1pt, topsep=1pt}

\newcommand{\toolfont}[1]{\textsc{#1}\xspace}
\newcommand{\madminer}[1]{\textsc{MadMiner}\xspace}

\newcommand{\diff}{\mathrm{d}}

\newcommand{\matel}[1]{\mathcal{M}_\text{#1}}

\newcommand{\ord}[1]{\mathcal{O}\left(#1\right)}

\newcommand{\pder}[2]{\frac {\partial #1} {\partial #2}}

\newcommand{\abc}{\textsc{ABC}\xspace}
\newcommand{\carl}{\textsc{Carl}\xspace}
\newcommand{\mem}{\textsc{MEM}\xspace}

\newcommand{\rolr}{\textsc{Rolr}\xspace}
\newcommand{\sally}{\textsc{Sally}\xspace}
\newcommand{\sallino}{\textsc{Sallino}\xspace}
\newcommand{\cascal}{\textsc{Cascal}\xspace}
\newcommand{\rascal}{\textsc{Rascal}\xspace}
\newcommand{\scandal}{\textsc{Scandal}\xspace}
\newcommand{\alice}{\textsc{Alice}\xspace}
\newcommand{\alices}{\textsc{Alices}\xspace}

\newcommand{\thetaref}{\theta_{\text{ref}}}

\newlength{\hhatheight}

\DeclareMathOperator*{\argmin}{arg\,min}

\newcolumntype{R}{>{\raggedleft\arraybackslash}X}%
\newcolumntype{L}{>{\raggedright\arraybackslash}X}%

\newcommand{\be}{\begin{equation} \begin{aligned}}
\newcommand{\ee}{\end{aligned} \end{equation}}

\begin{document}

\count\footins = 1000 


\title{Effective LHC measurements with matrix elements and machine learning}


\author{J.~Brehmer${}^1$, K.~Cranmer${}^1$, I.~Espejo${}^1$, F.~Kling${}^2$,
G.~Louppe${}^3$, and J.~Pavez${}^4$}

\address{${}^1$~New York University, USA;
${}^2$~University of California, Irvine, USA;
${}^3$~University of Li\`{e}ge, Belgium;
${}^4$~Federico Santa Mar\'ia Technical University,  Chile}

\ead{johann.brehmer@nyu.edu}

\begin{abstract}
  One major challenge for the legacy measurements at the LHC is that the likelihood function is not tractable when the collected data is high-dimensional and the detector response has to be modeled. We review how different analysis strategies solve this issue, including the traditional histogram approach used in most particle physics analyses, the Matrix Element Method, Optimal Observables, and modern techniques based on neural density estimation. We then discuss powerful new inference methods that use a combination of matrix element information and machine learning to accurately estimate the likelihood function. The \madminer{} package automates all necessary data-processing steps. In first studies we find that these new techniques have the potential to substantially improve the sensitivity of the LHC legacy measurements.
\end{abstract}

\section{LHC measurements as a likelihood-free inference problem}

The legacy measurements of the Large Hadron Collider (LHC) experiments aim to
put precise constraints on indirect effects of physics beyond the Standard
Model, for instance parameterized in an effective field theory
(EFT)~\cite{Buchmuller:1985jz}. This requires a careful analysis of
high-dimensional data. The relation between data $x$ and physics parameters
$\theta$ is most fundamentally described by the likelihood function or
normalized fully differential cross section $p(x|\theta) = {\diff\sigma(x |
\theta)} / {\sigma(\theta)}$. In fact, this likelihood function is the basis for
most established statistical methods in high-energy physics, including maximum
likelihood estimation, hypothesis testing, and exclusion limits based on the
profile likelihood ratio~\cite{Cowan:2010js}.

Typically, LHC processes are most accurately described by a suite of complex
computer simulations that describe parton density functions, hard process,
parton shower, hadronization, detector response, sensor readout, and
construction of observables with impressive precision. These tools take values
of the parameters $\theta$ as input and use Monte-Carlo techniques to sample
from the many different ways in which an event can develop, ultimately leading
to simulated samples of observations $x \sim p(x|\theta)$. The likelihood
that these simulators implicitly define can be symbolically written
as~\cite{Brehmer:2018kdj, Brehmer:2018eca}
\begin{equation}
  p(x|\theta) = \int\!\diff z_d \; \int\!\diff z_s \; \int\!\diff z_p \;
  \underbrace{p(x|z_s) \, p(z_s|z_p) \, p(z_p|\theta)}_{p(x,z|\theta)} \,,
  \label{eq:latent}
\end{equation}
where $z_d$ are the variables characterizing the detector interactions in one
simulated event, $z_s$ describes the parton shower and hadronization, and $z_p$
are the properties of the elementary particles in the hard interaction
(four-momenta, helicities, charges, and flavours). These latent variables form
an extremely high-dimensional space: with state-of-the-art simulators including
\toolfont{Geant4}~\cite{Agostinelli:2002hh} for the detector simulation, one
simulated event can easily involve tens of millions of random numbers!
Explicitly calculating the integral over this huge space is clearly impossible:
the likelihood function $p(x|\theta)$ is intractable.

This is not a problem unique to particle physics. Phenomena that are modeled by
a forward simulation that does not admit a tractable likelihood are common in
fields as diverse as cosmology, epidemiology, and systems biology. This has
given rise to the development of many different methods of ``likelihood-free
inference'' that allow us to constrain parameter values even without a
computable likelihood function. Phrasing particle physics measurements in this
language allows us to tap into recent developments in these fields as well as in
statistics and computer science.

\section{Established methods}

We will now briefly review six established types of inference methods that can
be used for particle physics measurements.

\subsection{Histograms of summary statistics}

Typically, not all the tens or hundreds of observables that can be calculated
for an event collision are equally informative on a given physics question.
Often it is enough to analyze one or two summary statistics $v(x)$, hand-picked
kinematic variables such as reconstructed invariant masses, momenta, or angles.
The likelihood function $p(v(x) | \theta)$ in the space of these summary
statistics can then be computed with simple density estimation techniques such
as one-dimensional or two-dimensional histograms, kernel density estimation
techniques, or Gaussian processes, and used instead of the likelihood function
of the high-dimensional event data. This approach discards any information in
the other phase-space directions.

This is by far the most common inference technique in high-energy physics. It is
fast and transparent. The disadvantage is, of course, that choosing the summary
statistics is a difficult and problem-specific task. While there are obvious
candidates for some problems, like the invariant mass in the case of the search
for narrow resonances, the indirect effects of EFT operators are typically not
captured well by any single observable~\cite{Brehmer:2016nyr, Brehmer:2017lrt}.
Histograms suffer from the ``curse of dimensionality'': since the required
number of samples grows exponentially with the dimension of $v$, they do not
scale to more than a few observables.

\subsection{Approximate Bayesian Computation}

Approximate Bayesian Computation (ABC)~\cite{rubin1984, beaumont2002approximate,
Alsing:2018eau, Charnock:2018ogm} is a family of Bayesian sampling techniques
for likelihood-free inference. Rather than providing a tractable surrogate of
the likelihood, ABC returns a set of parameter points $\theta \sim
\hat{p}(\theta|x)$ sampled from an approximate version of the posterior. In its
simplest form, this is based on accepting or rejecting individual simulated
samples $x$ based on some notion of distance to the observed data and an
distance threshold $\epsilon$.

As in the histogram case, this requires compressing the data $x$ to
summary statistics $v(x)$, again discarding any information in other variables.
In addition, the posterior is only exact in the limit $\epsilon \to 0$, but in
this limit the number of required simulations explodes.

\subsection{Neural density estimation}

In recent years, several machine learning techniques have been developed that
train a neural network to estimate the likelihood $\hat{p}(x|\theta)$ based only
on samples from the simulator~\cite{2012arXiv1212.1479F, 2014arXiv1410.8516D,
2015arXiv150203509G, 2015arXiv150505770J, Cranmer:2016lzt, 2016arXiv160508803D,
NIPS2016_6084, 2016arXiv160206701P, 2016arXiv161110242D, 2016arXiv160502226U,
2016arXiv160903499V, 2016arXiv160605328V, 2016arXiv160106759V,
gutmann2017likelihood, 2017arXiv170208896T, 2017arXiv170707113L,
2017arXiv170507057P, 2018arXiv180400779H, 2018arXiv180507226P,
2018arXiv181001367G}. Two particularly promising classes are autoregressive
models, which model a high-dimensional probability distribution as a product of
one-dimensional conditionals, and normalizing flows, which model the
distribution of $x$ as a simple base density followed by several invertible
transformations.

These neural density estimation techniques scale well to high-dimensional
observables, so they do not require a compression to summary statistics.
However, these algorithms can be less transparent than other approaches,
requiring careful cross-checks. In addition, they are agnostic about the physics
processes they are modeling: to the neural network, it does not make a
difference whether it estimates the density of smiling cats in the space of all
images or the density of LHC collisions for a particular EFT hypothesis. Often a
large number of training samples is necessary for the network to an accurate
estimate the true likelihood.

\subsection{Likelihood ratios from classifiers}

Other machine learning techniques are based on training neural networks to
classify between samples drawn from a given parameter point $x \sim p(x|\theta)$
and samples drawn from a reference parameter point $x \sim p(x|\thetaref)$. If
trained succesfully, the classifier decision function can be converted into an
estimator $\hat{r}(x|\theta)$ of the likelihood ratio
\begin{equation}
  r(x|\theta) = \frac {p(x|\theta)} {p(x|\thetaref)} \,.
\end{equation}
In practice, this is just as powerful as an estimator for the likelihood itself.
This ``likelhood ratio trick'' is widely appreciated~\cite{2014arXiv1406.2661G,
2016arXiv161003483M} and can be made much more accurate by adding a calibration
stage, resulting in the \carl inference technique~\cite{Cranmer:2015bka}.

Similar to the neural density estimation techniques, this approach is
well-suited to high-dimensional data and does not require choosing summary
statistics, at the cost of some (perceived) intransparency and a potentially
large number of required training samples.

\subsection{The Matrix Element Method}

An important insight is that while the integral in Eq.~\eqref{eq:latent} is
intractable, some parts of the integrand can in fact be calculated. In
particular, the parton-level likelihood function $p(z_p | \theta)$ is given by a
combination of phase-space factors, parton density functions, and matrix
elements, all of which we can compute. On the other hand, in many processes the
combined effect of parton shower and detector response is just a ``smearing'' of
the true parton-level momenta $z_p$ into the observed reconstructed particle
momenta $x$. If these effects can be approximated with a simple tractable
density $\hat{p}(x|z_p)$, the ``transfer function'', the likelihood is
approximately
\begin{equation}
  \hat{p}(x|\theta) = \int\!\diff z_p \; \hat{p}(x | z_p) \, p(z_p | \theta)
  \sim \frac 1 {\sigma(\theta)} \;
  \int\!\diff z_p \; \hat{p}(x | z_p) \, |\matel{}(z_p | \theta)|^2 \,.
  \label{eq:mem}
\end{equation}
In the last part we have left out parton densities as well as phase-space
factors for simplicity. This integral is much lower-dimensional than the one in
Eq.~\eqref{eq:latent}, and we can indeed often calculate this approximate
likelihood function! In the simplest case, the observed particle momenta are
identified with the parton-level momenta, the transfer function becomes
$\hat{p}(x|z_p) = \prod_i \delta^{4}(x_i - z_{p\,i})$, and the integration is
trivial.

This is the essential idea behind the Matrix Element Method
(MEM)~\cite{Kondo:1988yd, Abazov:2004cs, Artoisenet:2008zz, Gao:2010qx,
Alwall:2010cq, Bolognesi:2012mm, Avery:2012um, Andersen:2012kn, Campbell:2013hz,
Artoisenet:2013vfa, Gainer:2013iya, Schouten:2014yza, Martini:2015fsa,
Gritsan:2016hjl, Martini:2017ydu}. Recently, this approach has been extended to
include an explicit calculation of leading parton-shower effects (``shower
deconstruction'', ``event deconstruction'')~\cite{Soper:2011cr, Soper:2012pb,
Soper:2014rya, Englert:2015dlp}.

Unlike the neural network methods discussed above, the MEM explicitly uses our
knowledge about the physics structure of the hard process and does not rely on
the correct training of neural networks. However, there are two significant
downsides. The approximation of shower and detector effects with the transfer
function $\hat{p}(x | z_p)$ is not always accurate. Even if many resolution
effects can be modeled well with such an ad-hoc function, other physical effects
such as additional QCD radiation are very difficult or even impossible to
describe in this framework. At the same time, the integral over $z_p$ can still
be very expensive, and has to be calculated for every single event\,---\,which
can sum up to an immense computational cost when large numbers of events are
considered.

\subsection{Optimal Observables}

The matrix element information and the approximation in Eq.~\eqref{eq:mem} can
also be used to define observables
\begin{equation}
  O_i(x|\thetaref) = \pder {} {\theta_i}
    \log \left( \int\!\diff z_p \; \hat{p}(x|z_p) \, p(z_p | \theta) \right) \Biggr |_{\theta = \thetaref}
    = \frac {\int\!\diff z_p \; \hat{p}(x|z_p) \, \partial_i p(z_p | \thetaref)}
      {\int\!\diff z_p \; \hat{p}(x|z_p) \, p(z_p | \thetaref)}
  \label{eq:oo}
\end{equation}
with one component per theory parameter $\theta_i$. $\thetaref$ is a reference
parameter point, often chosen to be $\thetaref = 0$. In the literature this
approach is typically used with the identification of observed and true momenta
$\hat{p}(x|z_p) = \prod_i \delta^{4}(x_i - z_{p\,i})$, but the extension to
non-trivial transfer functions is straightforward. These observables can be used
like any kinematic variable. In particular, the likelihood in the space of the
$O_i(x)$ can be estimated using histograms or other density estimation
techniques.

In the approximation of Eq.~\eqref{eq:mem} and as long as only parameter points
close the $\thetaref$ are considered, the $O_i(x)$ are the sufficient
statistics: they fully characterize the high-dimensional event data $x$, and an
analysis based on $p(O_i(x)|\theta)$ will lead to the same conclusions as an
analysis of the intractable $p(x|\theta)$. The $O_i$ are therefore known as
Optimal Observables~\cite{Atwood:1991ka, Davier:1992nw, Diehl:1993br}.

The Optimal Observable approach shares the underlying approximation, strengths,
and disadvantages of the Matrix Element Method. While the immediate use of
matrix element information is beautiful, it requires stark approximations, and
any treatment of detector effects incurs a large computational cost for each
analyzed event.

\section{``Mining gold'': New inference techniques}

In a series of recent publications, a family of new techniques for
likelihood-free inference based on a combination of matrix element information
and machine learning was introduced~\cite{Brehmer:2018hga, Brehmer:2018kdj,
Brehmer:2018eca, Stoye:2018ovl}. They fall in two categories: a first class of
methods trains a neural network to estimate the full likelihood function, while
a second class is motivated by an expansion of the likelihood around a reference
parameter point and trains a neural network to provide optimal observables.

\subsection{Learning the high-dimensional likelihood}

The starting point for the new methods is the same as for the Matrix Element
Method: while the likelihood in Eq.~\eqref{eq:latent} is intractable because of
a high-dimensional integral, we can in fact compute the parton-level likelihood
function $p(z_p | \theta)$, given by a combination of phase-space factors,
parton density functions, and matrix elements. This means for each simulated
event we can also calculate the \emph{joint likelihood ratio}
\begin{equation}
  r(x,z|\theta) = \frac {p(x,z|\theta)} {p(x,z|\thetaref)}
  = \frac {p(z_p|\theta)} {p(z_p|\thetaref)}
  \sim \frac {|\matel{}|^2(z_p|\theta)} {|\matel{}|^2(z_p|\thetaref)} \; \frac {\sigma(\thetaref)} {\sigma(\theta)}
  \label{eq:joint_ratio}
\end{equation}
and the \emph{joint score}
\begin{equation}
  t(x,z|\theta) = \nabla_\theta \log p(x,z|\theta)
  = \frac {\nabla_\theta p(z_p|\theta)} {p(z_p|\theta)}
  \sim \frac {\nabla_\theta |\matel{}|^2(z_p|\theta)} {|\matel{}|^2(z_p|\theta)} - \frac {\nabla_\theta \sigma(\theta)} {\sigma(\theta)}
  \label{eq:joint_score}
\end{equation}
These two quantities define how much more or less likely one particular
evolution of an event (fixing all the latent variables $z$) would be if we
changes the theory parameters.

Why are these quantities useful? They are conditional on unobservable variables
$z$, most notably the parton-level momenta of the particles. But the key
insights behind the new methods is that the joint likelihood ratio and joint
score can be used to construct functionals $L[g(x, \theta)]$ that \emph{are
minimized by the true likelihood or likelihood ratio function!} In practice, we
can implement this minimization with machine learning: a neural network $g(x,
\theta)$ that takes as input the observables $x$ and the parameters
$\theta$~\citep[see][]{Cranmer:2015bka, Baldi:2016fzo} is trained by minimizing a
loss function that involves both the joint likelihood ratio $r(x,z|\theta)$ and
the joint score $t(x,z|\theta)$ via stochastic gradient descent (or some other
numerical optimizer). Assuming sufficient network capacity, efficient
minimization, and enough training samples, the network will then converge
towards the true likelihood function
\begin{equation}
  g(x, \theta) \to \argmin_g L_p[g(x, \theta)] = p(x | \theta)
\end{equation}
or, equivalently, the true likelihood ratio function
\begin{equation}
  g(x, \theta) \to \argmin_g L_r[g(x, \theta)] = r(x | \theta)\,!
\end{equation}
These loss functions thus allow us to turn tractable quantities based on the
matrix element into an estimator for the intractable likelihood function.

There are several loss functionals with this property, named with the acronyms
\rolr, \rascal, \cascal, \scandal, \alice, and \alices. They are individually
discussed and compared in Refs.~\cite{Brehmer:2018hga, Brehmer:2018kdj,
Brehmer:2018eca, Stoye:2018ovl}. In first experiments, the \alices
loss~\cite{Stoye:2018ovl} provides the best approximation of the likelihood
ratio, while the \scandal loss~\cite{Brehmer:2018hga} allows to directly
estimate the likelihood function and uses state-of-the-art neural density
estimation techniques in combination with the matrix element information.

Once a neural network is trained to estimate the likelihood (ratio) function,
established statistical techniques such as profile likelihood ratio
tests~\cite{Cowan:2010js} can be used to construct confidence limits in the
parameter space.

This approach can be seen as a generalization of the MEM technique that supports
state-of-the-art shower and detector simulations as opposed to simple transfer
functions. While it requires an upfront training phase, the evaluation of the
likelihood for each event is extremely fast (``amortized inference'').

\subsection{Learning locally optimal observables}

Rather than learning the full likelihood function, the joint score can also be
used to define statistically optimal observables. This approach is motivated by
an expansion of the log likelihood in theory space around a reference
parameter point $\thetaref$ (such as the SM):
\begin{equation}
  \log p(x|\theta) = \log p(x|\thetaref) + t(x|\thetaref) \cdot (\theta - \thetaref) + \ord{(\theta - \thetaref)^2}
\end{equation}
where we have introduced the \emph{score}
\begin{equation}
  t(x|\theta) = \nabla_\theta \log p(x|\theta) \,.
\end{equation}

As long as we are considering parameter points close enough to $\thetaref$,
we can neglect the higher orders, and the score vector fully characterizes the
likelihood function up to $\theta$-independent constants. In fact, in this local
approximation the likelihood is in the exponential family and the score
components are the sufficient statistics: for measuring $\theta$, knowing the
$t(x|\thetaref)$ is just as powerful as knowing the full likelihood
function~\cite{Brehmer:2018hga, Brehmer:2018kdj, Brehmer:2018eca,
Alsing:2017var, Alsing:2018eau}. The score at $\thetaref$ is thus a vector of
the statistically most powerful observables! Further away from $\thetaref$, the
higher-order terms become important, and the score loses this optimality
property.

Unfortunately, the score itself is defined through the intractable likelihood
function and cannot be calculated explicitly. But it is possible to compute the
joint score of Eq.~\eqref{eq:joint_score} for each simulated event. Similarly to
the approach discussed above, we can train a neural network $g(x)$ on a suitable
loss function and show that it will converge to
\begin{equation}
  g(x) \to \argmin_g L_t[g(x)] = t(x | \theta)\,.
\end{equation}
In this way, we can train a neural network to define the most powerful
observables. In a next step, the likelihood can be determined for instance with
simple histograms of the score components. This is the basic idea behind the
\sally and \sallino inference techniques of Refs.~\cite{Brehmer:2018hga,
Brehmer:2018kdj, Brehmer:2018eca}.

This approach is particularly robust and requires only minor changes to
established analysis pipelines. Note that the score vector is almost the same as
the Optimal Observables $O_i$ of Eq.~\eqref{eq:oo}, but replaces the
parton-level (or transfer-function) approximation with a neural network
estimation of the full statistical model, including state-of-the-art shower and
detector simulations.

\subsection{Discussion}

\begin{table}
\centering
\scriptsize
\begin{tabular*}{0.97\textwidth}{@{\extracolsep{\fill}} l c c@{~}c@{~}c@{~}c c c}
\toprule
\multirow{2}{*}{Method} & \multirow{2}{*}{Estimates} & \multicolumn{4}{c}{Approximations}
& \multirow{2}{*}{$|\mathcal{M}|^2$} & \multirow{2}{*}{Comp.~cost} \\
\cmidrule{3-6}
& & summaries & PL/TF & local & functional \\
\midrule
Histograms of observables & $\hat{p}(v(x)|\theta)$ & $\checkmark$ & & & binning & & low \\
Approximate Bayesian Computation & $\theta \sim p(\theta|x)$ & $\checkmark$ & & & $\epsilon$-kernel & & high (small $\epsilon$) \\
Neural density estimation & $\hat{p}(x|\theta)$ & & & & NN & & amortized \\
\carl & $\hat{r}(x|\theta)$ & & & & NN & & amortized \\
Matrix Element Method & $\hat{p}(x|\theta)$ & & $\checkmark$ & & integral & $\checkmark$ & high (TF) \\
Optimal Observables & $\hat{t}(x)$ & & $\checkmark$ & $\checkmark$ & integral & $\checkmark$ & high (TF) \\
\midrule
\scandal & $\hat{p}(x|\theta)$ & & & & NN & $\checkmark$ & amortized \\
\alice, \alices, \cascal, \rascal, \rolr & $\hat{r}(x|\theta)$ & & & & NN & $\checkmark$ & amortized \\
\sally, \sallino & $\hat{t}(x)$ & & & $\checkmark$ & NN & $\checkmark$ & amortized \\
\bottomrule
\end{tabular*}
\caption{Classes of established (top part) and novel (bottom half) inference
techniques. We classify them by the key quantity that is estimated in the
different approaches, by whether they rely on the choice of summary statistics,
are based on a parton-level or transfer-function approximation (``PL/TF''),
whether their optimality depends on a local approximation (``local''), by whether
they use any other functional approximations such as a histogram binning or a neural
network (``NN''), whether they leverage matrix-element information
(``$|\mathcal{M}|^2$''), and by the computational evaluation cost.}
\label{tbl:comparison}
\end{table}

In Table~\ref{tbl:comparison} we compare the different approaches roughly.
First, all techniques rely on some approximations or simplifications to make the
likelihood tractable. For the traditional histogram and \abc techniques, this is
the restriction to one or a few summary statistics, which potentially throws
away a substantial amount of information. For the matrix element method and
Optimal Observables, neglecting or approximating the shower and detector
response plays this role, which is expected to have a larger impact the less
clean a final state is. For neural density estimation techniques, likelihood
ratio estimation based on classifiers (\carl), or the new techniques presented
in this section, the key approximation is the reliance on neural networks to
minimize a functional. Several diagnostic tools and calibration procedures have
been proposed~\cite{Cranmer:2015bka, Brehmer:2018kdj} that make the network
predictions more trustworthy and can guarantee statistically correct (if
potentially suboptimal) limits. In addition, both the traditional Optimal
Observable method and the new \sally and \sallino techniques are based on a
local approximation: the summary statistics they define are statistically
optimal only within a small region in parameter space.

Second, the different methods have very different computational costs and scale
differently to high-dimensional problems. The classical histogram approach can
be relatively cheap: if samples are generated for every tested parameter point
on a grid, the number of required simulator runs scales as $ e^{n_\theta}$,
where $n_\theta$ is the dimension of the parameter space. But a morphing
technique can typically be used to make the number of required runs scale as
$n_\theta^2$ or $n_\theta^4$~\cite{ATLAS:morphing, Brehmer:2018kdj}. \abc can be
expensive for small $\epsilon$ or when many parameter points from the posterior
are required. Since they are based on explicitly calculating integrals over
latent variables, the Matrix Element Method and Optimal Observable approaches
scale differently from the other methods: the run time scales approximately
exponential in the number of unobservable directions in phase space $n_z$. In
terms of $n_\theta$, the calculation time scales as $n_\theta$ in the case of
Optimal Observables and $e^{n_\theta}$ in the case of the \mem evaluated on a
parameter grid. A major disadvantage of these methods is that the calculation of
the integral has to be repeated for every new event, resulting in an overall
$e^{n_z} n_\theta n_x$ (OO) or $e^{n_z} e^{n_\theta} n_x$ (\mem) scaling.

The methods based on machine learning, on the other hand, allow for
\emph{amortized inference}: after an upfront training phase, the evaluation of
the approximate likelihood, likelihood ratio, or score is extremely fast. The
key question is therefore how much training data is required for an accurately
trained model. For the methods that estimate the likelihood or likelihood ratio,
this scaling depends on the specific problem: in the worst case, when
distributions vary strongly over the parameter space, the number of Monte-Carlo
samples is expected to scale as $e^{n_\theta}$. In practice, distributions
change smoothly over parameter space, and parameterized neural networks can
discover these patterns with little data. Once the network is trained, inference
is fast (unless large toy measurement samples are required for calibration,
which scales with $e^{n_\theta}$ again). How much the augmented data improves
sample efficiency depends again on the specific problem: a larger effect of
shower and detector on observables increases the variance of the joint
likelihood ratio and joint score around the true likelihood ratio and true score
and finally the number of required samples.

In the \sally\,/\,\sallino approach, the networks do not have to learn a
function of both $x$ and $\theta$, but rather an $n_\theta$-dimensional vector
as a function of $x$\,---\,this turns out to be a substantially simpler problem
and requires much less training data. The main difference between them is in the
number of simulations required to calculate the likelihood for a single
parameter point. \sally requires filling an $n_\theta$-dimensional histogram,
for which the number of required samples scales like $e^{n_\theta}$, while
$\sallino$ is based on one-dimensional histograms.

\section{MadMiner: A sustainable software environment}

We are developing \madminer{}~\cite{madminer_paper}, a Python package that
automates all steps of the new inference techniques from the running of the
Monte-Carlo simulations, extracting the augmented data, training neural network
with suitable loss functions, and calculating expected and observed limits on
parameters. The current version v0.4.0 wraps around
\toolfont{MadGraph5\_aMC}~\cite{Alwall:2014hca},
\toolfont{Pythia~8}~\cite{Sjostrand:2007gs}, and
\toolfont{Delphes~3}~\cite{deFavereau:2013fsa}, providing all tools for a
phenomenological analysis. Its modular interface allows the extension to
state-of-the-art tools used by the experimental collaborations; such an
extension would mostly require book-keeping of event weights.

\madminer{} is open source, available on GitHub~\cite{MadMiner_repo}, and can be
installed with a simple \pyth{pip install madminer}. Its documentation is
published on ReadTheDocs~\cite{MadMiner_docs}. We provide interactive
tutorials~\cite{MadMiner_repo}, a Docker container with a working software
environment~\cite{MadMiner_docker}, and deploy \madminer{} with a reusable
\toolfont{Reana} workflow~\cite{MadMiner_reana}.

\section{In the wild}

The new inference techniques are used in several real-life projects. The
original publications in Refs.~\cite{Brehmer:2018hga, Brehmer:2018kdj,
Brehmer:2018eca, Stoye:2018ovl} contained a proof-of-concept EFT analysis of
Higgs production in weak boson fusion (WBF) in the four-lepton decay mode. In a
somewhat simplified setup, the new methods could substantially improve the
expected sensitivity to two dimension-six operators compared to histogram-based
analyses. They also required more than two orders of magnitude less training
samples than the \carl method.

After these encouraging first results, the new methods and \madminer{} are now
being used in several realistic analyses, including $Wh$~production~\cite{wh},
$W\gamma$~production~\cite{wgamma}, and $t\bar{t}h$
production~\cite{madminer_paper}. Other ongoing projects include a comparison of
the new techniques with the Matrix Element Method and applications of the new
techniques to problems in other fields.

\section{Conclusions}

The LHC legacy measurements will require picking out (or excluding) subtle
signatures in high-dimensional data to put limits on an also high-dimensional
parameter space. While Monte-Carlo simulations provide an excellent description
of this process, they do not allow us to explicitly calculate the corresponding
likelihood function. We have reviewed how different analysis strategies address
this issue, including the traditional histogram approach, the Matrix Element
Method, Optimal Observables, and methods based on machine learning.

We then discussed a new family of multivariate inference techniques that combine
matrix element information with machine learning. This new paradigm brings
together the strengths of different existing methods: the methods do not require
the choice of a summary statistic, utilize matrix element information
efficiently, but unlike the Matrix Element Method or Optimal Observables they
support state-of-the-art shower and detector simulations without approximations
on the underlying physics. After an upfront training phase, they can also be
evaluated very fast.

The new Python package \madminer{} automates all steps of the analysis chain. It
currently supports all tools of a typical phenomenological
analysis\,---\,\toolfont{MadGraph5\_aMC}, \toolfont{Pythia~8},
\toolfont{Delphes~3}\,---\,but can be scaled up to experimental analyses.

A first analysis of EFT operators in WBF Higgs production showed that the new
techniques led to stronger bounds with less training data compared to
established methods. Several studies of other channels are now underway. If they
confirm the initial results, these new techniques have the potential to
substantially improve the precision of the LHC legacy measurements.

\subsection*{Acknowledgements}

We would like to thank Zubair Bhatti, Pablo de Castro, Lukas Heinrich, and
Samuel Homiller for great discussions, and we are grateful to the ACAT
organizers for a wonderful workshop. This work was supported by the National
Science Foundation under the awards ACI-1450310, OAC-1836650, and OAC-1841471.
It was also supported through the NYU IT High Performance Computing resources,
services, and staff expertise. JB, KC, and GL are grateful for the support of
the Moore-Sloan data science environment at NYU. KC is also supported through
the NSF grant PHY-1505463, FK is supported by NSF grant PHY-1620638,
while JP is partially supported by the Scientific and Technological Center of
Valpara\'{i}so (CCTVal) under Fondecyt grant BASAL FB0821.

\newcommand{\newblock}{}
\bibliographystyle{jthesis}
\bibliography{references}

\end{document}